# High-$Q$ Resonances Governed by the Quasi-Bound States in the Continuum in All-Dielectric Metasurfaces

Cizhe Fang [1], Qiyu Yang [1], Qingchen Yuan [2], Xuetao Gan [2]*, Jianlin Zhao [2], Yao Shao [3], Yan Liu [1]*, Genquan Han [1], Yue Hao [1]

[1] State Key Discipline Laboratory of Wide Band Gap Semiconductor Technology, Shaanxi Joint Key Laboratory of Graphene, School of Microelectronics, Xidian University, Xi'an, 710071, China;

[2] MOE Key Laboratory of Material Physics and Chemistry under Extraordinary Conditions, and Shaanxi Key Laboratory of Optical Information Technology, School of Physical Science and Technology, Northwestern Polytechnical University, Xi'an, 710129, China

[3] Shanghai Energy Internet Research Institute of State Grid, 251 Libing Road, Pudong New Area, Shanghai, 201210, China.

**Abstract**: The realization of high-$Q$ resonances in a silicon metasurface with various broken-symmetry blocks is reported. Theoretical analysis reveals that the sharp resonances in the metasurfaces originate from symmetry-protected bound in the continuum (BIC) and the magnetic dipole dominates these peculiar states. A smaller size of the defect in the broken-symmetry block gives rise to the resonance with a larger $Q$ factor. Importantly, this relationship can be tuned by changing the structural parameter, resulting from the modulation of the topological configuration of BICs. Consequently, a $Q$ factor of more than 3,000 can be easily achieved by optimizing dimensions of the nanostructure. At this sharp resonance, the intensity of the third harmonic generation signal in the patterned structure can be 368 times larger than that of the flat silicon film. The proposed strategy and underlying theory can open up new avenues to realize ultrasharp resonances, which may promote the development of the potential meta-devices for nonlinearity, lasing action, and sensing.

## Introduction

In the last decades, research into optical metasurfaces[1,2] has been intensified because of their multifunctionality, which can be employed for many applications in photonics,[3-11] chemistry and biosensing.[12,13] Metasurfaces are ultrathin planar arrangements consisting of subwavelength spaced scatterers with different shapes and sizes. They can modify the amplitude, phase, and polarization of propagating optical waves upon transmission or reflection.[14-18] However, due to the Ohmic loss, the conventional plasmonic metasurfaces[19] suffer from high losses and heat dissipation, which hinders their application in various functional nanoscale devices,[20-22] especially low-loss meta-devices. To circumvent this constraint, all-dielectric metasurfaces are of paramount interest in recent years due to their high efficiency.[22,23] For high-index dielectric material, individual nanoparticles can possess electric and magnetic Mie-type resonances,[22,24,25] leading to the electromagnetic field confinement and multiple interferences inside the particles.[26] However, nanoparticles made of conventional optical materials (e.g. silicon (Si)) can only support a relatively low quality factor ($Q$ factor).[27] This broad spectral response limits their applicability to the devices requiring sharp spectral features such as lasing and sensing.[28,29] A recently emerged concept of BIC provides a new solution to overcome this problem.[27-30]

BICs represent a wave phenomenon of modes, which have the energy lying in the delocalized states inside the continuum. This concept firstly appears in the quantum mechanics,[30,31] and then BICs are observed in acoustics, optics, microwaves, and so on.[32-34] A true BIC is a mathematical abstraction with infinite $Q$ factor and the access to radiative channels is completely forbidden. In fact, for a practical device (i.e. finite-size structure), all possible BICs turn into the leaky mode with a high $Q$ factor,[35] also known as a quasi-BIC mode. The recent work also demonstrates that there exists a direct link between this leaky mode and Fano resonances.[35] So far, tremendous work has been explored to distinguish different mechanisms of BICs.[32-37] Among them, symmetry-protected BICs show a $Q$ factor dependence on the asymmetry parameter $\alpha$,[38] which provides a potential strategy to design all-dielectric metasurfaces supporting high-$Q$ resonance.

In this paper, we introduce the defects in the unit cell of the periodic metasurfaces based on Si. The quasi-BIC modes are emerged, which manifest themselves as Fano resonances. By employing the finite element method (FEM), the physical mechanism is studied. The theoretical results also show the relationship between the asymmetry parameter and the $Q$ factor, which are further confirmed by the experiment. Moreover, it is proved that the high-$Q$ resonance can dramatically enhance the intensity of the third harmonic generation (THG) signal. We believe that these results can be not only useful for improving the nonlinear conversion processes,[39,40] but also for lasing and sensing applications.

## Device Design and Theoretical Analysis

First, we consider a design of the periodic structure with in-plane symmetry to support BIC modes (see the inset of Fig. 1(a)). The mode properties are studied by utilizing the FEM. For simplicity, a single unit in free space is simulated in $xy$ planes with periodic boundary conditions while perfectly matched layers are used along the $z$-axis.[41] The plane waves with $x$-polarization are incident from the $z$-axis and the refractive index of silicon is set as 3.48. The scaling parameter, the side length of the square block, the thickness of the block, the period, and incident angle are denoted by $s$, $l$, $t$, $p$, and $\theta$, respectively. In all simulations, $p$ and $t$ are fixed at 0.6 μm and 0.23 μm, respectively. Fig. 1(a) describes the $Q$ factor of the mode for different $\theta$ and $l$. One can see that the mode has an infinite $Q$ factor (more than $10^8$) at the normal incidence. With the increase in $\theta$, the $Q$ factor becomes finite and drops rapidly. By contrast, this mode is insensitive to geometrical change (i.e. the side length, $l$) and preserves high-$Q$ mode (i.e. ~$10^8$) with the change in $l$, indicating that it is a symmetry-protected BIC mode.[28,36] For better understanding, the transmission spectrum dependence on angular incidence is plotted in Fig. 1(b). It can be seen that the resonance (manifested as dips in the transmission) narrows and then vanishes when the incident angle approaches 0°, accompanied by the blue shift in the resonant wavelength. It corresponds to the decline in radiation loss and the emergence of the symmetry-protected BIC. To gain a deeper insight into the physics of this decline, the magnetic field profiles at $\theta$=0° are displayed in the inset of Fig. 1(b). The electric field presents distinct $z$-directed magnetic dipole (MD) features in one unit cell, manifesting themselves as circular currents.[42] Simultaneously, the intensity of the magnetic field decreases rapidly once away from the block since the intensity of the magnetic field at $z$=0.345 μm nearly decreases by an order of magnitude compared with that at $z$=0 μm. These modes do not radiate in the vertical direction along the $z$-axis because the normal radiative decay is compensated by driving terms arising from the local field at the position of the resonator.[43,44] After introducing the defect, this compensation is broken and the ideal BIC transforms into the leaky mode with the sharp resonance. The radiation powers of the MD moments are the dominant factor for these resonances. As can be noticed from the top two insets of Fig. 1(c), a distinct MD can be observed in the block with broken in-plane symmetry. To interpret the influence of the detect size on the $Q$ factor, the different defects are introduced in the blocks (see the bottom inset of Fig. 1(c). The size of the defect is only controlled by $a$ since $b$ is fixed at 0.1 μm. As shown in Fig. 1(c), with the increase in the defect size, the $Q$ factor decreases sharply. When the defect size continues to increase (i.e. $a$>0.2 μm), the $Q$ factor begins to increase. The $Q$ factor does not monotonically decrease with the increase in the size of the defect, which is not consistent with the previous report.[38] We ascribe this phenomenon to the existence of another BIC mode at $a$=0.4 μm (see the bottom left inset in Fig. 1(d)) and eigenvalue simulations are carried out to verify this speculation. Fig. 1(d) shows that the mode is also extremely sensitive to $\theta$ as its quality factor drops sharply for a small $\theta$. However, the $Q$ factor is independent of $l$ and remains high. Meanwhile, the electromagnetic field is perfectly localized in the block (see the bottom right inset of Fig. 1(d)), implying that these high-$Q$ resonances also originate from the symmetry-protected BIC mode. When $\theta$ continues to increase, the $Q$ factor begins to increase along with $\theta$. It is believed that this phenomenon results from the existence of a resonance-trapped mode at $\theta$=19° (indicated by the red dotted circle), which also shows a large $Q$ factor. It is worth noting that compared to the symmetry-protected BIC, the resonance-trapped BIC is relatively less sensitive to $\theta$ and thus remains a high $Q$ factor at around $\theta$=19°.[28,36] It occurs in the metasurface due to the destructive interference[34, 36] and it is sensitive to geometric dimensions (i.e. $l$). To prove it, the $Q$ factor is modeled as a function of $l$ at $\theta$=19° (see the top inset in Fig. 1(d)). Compared with the influence of $\theta$, the $Q$ factor decreases more dramatically once $l$ leaves the optimal value. It is also different from the insensitivity of the $Q$ factor to $l$ in the symmetry-protected mode. The further study of this mode is beyond the scope of this work.

Owing to the fact that there exist two different BIC modes (i.e. $a$=0 μm and $a$=0.4 μm), we define the asymmetry parameter $\alpha$ for both cases as seen in the inset of Fig. 2(a). It can be seen that for small values of $\alpha$, a smaller asymmetry parameter corresponds to a larger $Q$ factor. However, the behavior of the $Q$ factor on $\alpha$ is different, where the slopes of the fitting lines are -0.61 and -1.8 for definition 1 and definition 2, respectively. This phenomenon can be explained by the topological configuration of BICs. Taking into account that the leaky mode arises when the BIC mode couples to far-field radiation, we focus on the model supporting the ideal BIC mode (i.e. $\alpha$=0). The models for definitions 1 and 2 are shown in the inset of Fig. 2(b), corresponding to the square block (see the inset of Fig. 1(a)) and rectangle block (see the bottom left inset of Fig. 1(d)), respectively. To verify this, the $Q$ factor distributions around the $\Gamma$ point are calculated and plotted for both scenarios. A comparison is shown in Fig. 2(b), where BICs are generated at the $\Gamma$ point for both cases. It arises because coupling constants with all radiating waves vanish due to symmetry. At this point, the light becomes perfectly confined in the metasurfaces, as is evident from the $Q$ factors and field profiles (see the magnetic field distributions shown in Figs. 1(b) and 1(d)). For the case of $a$=0.4 μm, there are two additional BICs (indicated by white circles) on the $k_y$ axis, which contributes to the variation in the $Q$ factor distribution along $k_y$ axis. The BIC at off-$\Gamma$ corresponds to the resonance-trapped mode in Fig. 1(d). Since the BICs behave like a topological defect, it causes the difference in the topological configuration between both structures. It is known that the topological configuration of BICs can control radiative losses of nearby resonances.[45] That indicates the reason for different decay rates of the $Q$ factor in two structures. As a result, the corresponding leaky modes (i.e. $\alpha \neq 0$) are affected and it finally leads to the different dependence of $Q$ factor on $\alpha$, as shown in Fig. 2(a). Moreover, in accordance with [46], the generation, evolution, and annihilation of BICs can be governed by varying structural parameters, which provides theoretical support to adjust the relationship between $Q$ factor and $\alpha$. Above all, the modulation of the topological configuration of BICs is also achievable in the metasurface. Therefore, this mechanism provides an important and simple tool to engineer the topological configurations and thus the behavior of the $Q$ factor on $\alpha$, which provides a platform for developing the realistic BIC devices upon metasurface.

To confirm such theoretical findings, metasurfaces with different dimensions using electron-beam lithography (EBL), followed by inductively coupled plasma (ICP) etching, were fabricated, as shown in Fig. 2(c). Fig. 2(d) presents the scanning electron microscope (SEM) images of the fabricated device. It should be pointed out that the dimensions of the blocks are always smaller than their design values, resulting from the fabrication tolerance, and it leads to a blue shift in resonant wavelength.

## Results and discussion

In this section, we focus on the discussion of measured results. The linear optical characteristics of the fabricated samples were measured using a home-built setup, where a continuous-wave (CW) laser was focused onto the sample, and the transmitted or reflected light was monitored using a spectrometer. Fig. 3(a) depicts the transmission spectra for different $s$. One can observe that transmission spectra show sharp resonances and the resonance positions cover the spectral range from 1.2 μm to 1.5 μm. The $s$-dependence on resonant wavelength is also valid beyond this spectrum range. It is noted the measured resonance appears as a peak in the transmission spectrum, different from the simulation results. We attributed this phenomenon to the result of the combined action of the Fano resonance and the substrate. It is known to us that the quasi-BIC manifests as itself the Fano resonance in the transmission. For the prepared metasurface, there exists a substrate. On the one hand, the substrate introduces the perturbation (i.e. asymmetry along the $z$-axis). At around the resonant wavelength, the mode leaks to the substrate (i.e. the incident direction), which increases the transmission. For better understanding, the magnetic field distribution in the $yz$ plane is plotted in the inset of Fig. 3(a). It is clear that at the resonant wavelength, there exists more leakage into the substrate. On the other hand, the substrate suppresses the transmitted light. As a result, the “dip” is suppressed and the Fano resonance exhibits as a “peak”. We also perform the polarization dependence of transmission and reflection spectra. As the polar plots are shown in Fig. 3(b), both transmission and reflection spectra show a distinct polarization dependence, consistent with the results of the simulation. Now, we turn toward the influence of the defect size on the $Q$ factor. Due to the process tolerance, only the proposed metasurfaces with $a$ ranging from 0.08 μm to 0.3 μm can meet the design requirements. The insets of Fig. 3(c) show that the shapes of the unit cells are greatly changed at $a$ =0.02 μm and 0.32 μm. For comparison, $Q$ factors for identical structures are also calculated and shown in Fig. 3(c). We observe a very good agreement between the experimental and calculated results. The $Q$ factor decreases with the increase in $a$ at first and then increases along with $a$. Notably, there is a large difference between the experimental and calculated results. It will be explained in the next paragraph. As shown in the bottom figure in Fig. 3(c), the measured $Q$ factor is modeled as a function of $\alpha$ for two definitions. It can be seen that the $Q$ factor increases with a decrease in $\alpha$. More importantly, there indeed exist two different relationships (i.e. the slopes are -0.35 and -0.69) between the $Q$ factor and asymmetry parameter $\alpha$, which confirms the results of the eigenvalue analysis. It is noted that the slopes of the measured results are different from that of the calculated results, especially for definition 2. We attribute this difference to the non-ideal factor in the experiment. According to the definition, the structures remove and add the perturbation for definitions 1 and 2, respectively. For convenience, we designed the pattern by removing the perturbation in the middle of the structure (see the bottom inset of Fig. 1(c)). By changing $a$, both structures can be realized. In this case, for the same $s$, the structure holds a smaller $A$ for definition 2. It means that the introduced perturbation is more easily affected by the non-ideal factor in the preparation process.

Next, we aim to discuss the implementation of BIC-supporting metasurfaces. As shown in Fig. 3(c), the measured $Q$ factor is several times smaller than that calculated one. Four major factors that lead to the low-$Q$ resonance, which are listed as follows. First, as an experimental sample, surface cleanliness is inevitable. It can introduce an inherent loss caused by scattering,[47] which finally leads to the attenuation in the $Q$ factor. The second one is fabrication tolerance, which has a great impact on the shape and size of the unit cell. As shown in the bottom inset of Fig. 3(d), after the process optimization, the surface is cleaner and the edges of the blocks are smoother in the structure. Importantly, the $Q$ factor can be improved from 250 to 413. Through process optimization, it can be realized to minimize the influences of the surface cleanliness and process preparation. Another long-standing factor is the substrate. The existence of the substrate will generate a field penetrating, which turns a BIC mode into the resonant one.[47] To verify this, we calculated and plotted the evolution of the $Q$ factor and resonant wavelength as a function of the thickness of the substrate $t_{sub}$. From Fig. 3(d), we can see that the $Q$ factor rapidly drops with the existence of the substrate accompanied by the redshift in resonant wavelength. As the thickness continues to increase, the $Q$ factor becomes relatively insensitive to $t_{sub}$. Besides, the array size also plays a crucial role in achieving a sharp resonance. As depicted in Fig. 3(d), the prepared metasurface with larger array sizes shows a larger $Q$ factor for both structures. It can be explained that with the increase in the array size, a finite array of the unit cell becomes an infinite periodic structure, which corresponds to an ideal BIC with an infinite value of the $Q$ factor. For bettering understanding, we simulated the response of the different size arrays and their corresponding $Q$ factor (see Fig. 4). The magnitude of the magnetic field increases along with the array size and the innermost blocks of the array show larger field enhancement than that at the edge of the array. As we mentioned above, these high-$Q$ modes hardly radiate in the vertical direction because the normal radiative decay is compensated by driving terms arising from the local field at the position of the resonator. However, for a finite array, the unit cell at the edge can only be partially compensated, leading to extra leakage. With the increase in the array size, the proportion of the unit cell at the edge can be reduced, corresponding to a less leakage (i.e. larger $Q$ factor). Above all, the low-$Q$ measured results are caused by the existence of the substrate and the finite array size, which are inevitable in practical application at the nanoscale. Therefore, process optimization and reasonable design can be more effective solutions to achieve a high $Q$ factor.

As we mentioned above, an optimized metasurface was fabricated to improve the $Q$ factor. Due to the resolution of the spectrometer, a narrowband tunable continuous-wave (CW) laser was utilized, and the transmitted optical powers were monitored using a telecom-band photodiode. Fig. 5(a) illustrates the transmission spectrum at a wavelength of around 1.55 μm. The transmission peak is fitting by a Fano formula:[48]

$$F(\varepsilon)=\frac{(q+\varepsilon)^2+\gamma^2}{1+\varepsilon^2}, \quad (1)$$

$$\varepsilon=(\omega-\omega_0)/\gamma, \quad (2)$$

$$Q=\omega_0/2\gamma, \quad (3)$$

where $\varepsilon$ represents the scale of reduced energy, $q$ is the asymmetry factor, $\omega_0$ is the resonant frequency, and $\gamma$ represents the damping rate. By fitting the data with Eq. (1), we can obtain the fitting parameters of $\omega_0$=1547.85 and $\gamma$=0.219. A $Q$ factor as high as 3534 can be obtained, which can be further improved by introducing smaller defects. As one of the most promising applications,[49] we consider nonlinear effects in the proposed metasurface. For the nonlinear measurements, the metasurface was pumped by a pico-second pulsed laser, and the transmitted THG signals were monitored using a spectrometer. As we can see from Fig. 5(b), the intensity of THG signals in the proposed structure is 368 times larger than that in the flat Si film. This enhancement can be ascribed to high-$Q$ resonances and strong field confinement,[19,50] which are considered as the dominant factors in nonlinear optics instead of phase matching at the nanoscale.[51] On the one hand, for this nonlinear response, the intensities of the THG scale approximately with the factor $(Q/V)^3$,[52] where $V$ is the mode volume. An enhancement in $Q$ factor at the fundamental frequency can be further amplified in the third-order nonlinear interaction. On the other hand, strong light confinement means that the photons are localized in a small space for a long time, which can enhance various light-matter interactions. Also, the electric field distributions at both the resonance and the off-resonance are plotted in Fig. 5(c). As we can see, there is a significant attenuation in the intensity of the electric field once away from the resonant wavelength, corresponding to the decrease in the intensity of the THG. Fig. 5(d) compares the polarization dependences of both THG. The pronounced THG from the flat Si film (black line) has another polarization dependence (indicated by the black arrow). This polarization dependence results from the THG in the sapphire substrate. Noticing that this polarization dependence disappears in the proposed structure (red line), it is because Si blocks possess the strong electromagnetic field, resulting in the significant enhancement in the intensity of the THG. It provides further evidence of the improvement in the THG signal caused by the proposed structure. The nonlinear response can be further enhanced and tailored by varying the dimensions of the metasurfaces.

# 4 Conclusions

In summary, we have successfully designed and fabricated Si metasurfaces with high $Q$ resonances, which have elements of broken-symmetry blocks. The MD modes of the blocks contribute to these resonances. Both simulation and experiment results reveal that the $Q$ factor increases when the size of the defects decreases and this relationship can be altered by changing the dimensions of the structure. As an example of the applications of the high $Q$ resonance, high-efficiency THG is implemented from the silicon metasurface, which presents an enhancement factor of 368. The results provided by this work pave a way to manipulate BICs and realize high-$Q$ dynamic resonances, and they constitute a significant step towards the development of high-$Q$ resonant photonic applications such as lasers, sensors, and filters.

## Acknowledgements

The authors acknowledge support from the National Key Research and Development Project (Grant No. 2018YFB2200500, 2018YFB2202800) and the National Natural Science Foundation of China (Grant No. 61534004, 91964202, 61874081, 61851406, 91950119, and 61905196).

## Author contributions

All authors reviewed and revised the manuscript with valuable suggestions. Y. Liu and X. Gan proposed the original idea. C. Fang and Y. Shao completed the simulation and analysis. C. Fang, Q. Y, and Q. Yuan fabricated the samples and performed the measurements. J. Zhao, G. Han, and Y. Hao provided experiment instruments (including measurement and characterization).

## Competing interests

The authors declare no competing financial interests.

# Authors:

Ph.D. student Cizhe Fang

State Key Discipline Laboratory of Wide Band Gap Semiconductor Technology, Shaanxi Joint Key Laboratory of Graphene, School of Microelectronics, Xidian University, Xi’an, 710071, China

Prof. Yan Liu

State Key Discipline Laboratory of Wide Band Gap Semiconductor Technology, Shaanxi Joint Key Laboratory of Graphene, School of Microelectronics, Xidian University, Xi’an, 710071, China

E-mail: xdliuyan@xidian.edu.cn

Prof. Xuetao Gan

MOE Key Laboratory of Material Physics and Chemistry under Extraordinary Conditions, and Shaanxi Key Laboratory of Optical Information Technology, School of Physical Science and Technology, Northwestern Polytechnical University, Xi’an, 710129, China

E-mail: xuetaogan@nwpu.edu.cn

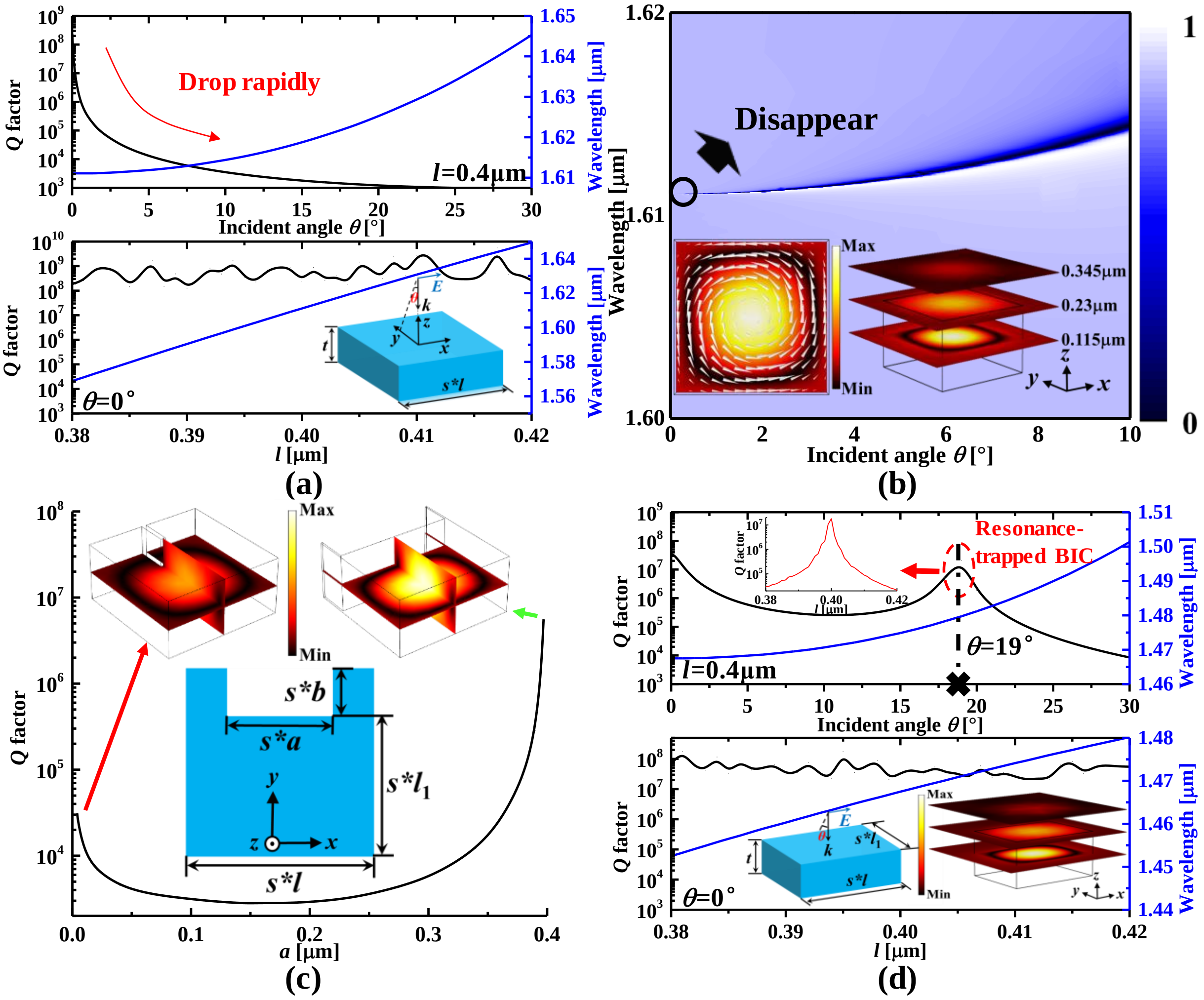

Fig. 1 | (a) Simulated $Q$ factor and resonant wavelength as a function of $\theta$ (top) and $l$ (bottom), respectively. Inset: the schematic view of the model (a square block with $l$=0.4 μm). (b) Simulated transmission spectrum as a function of the angle and wavelength. Inset: the magnetic field profiles under the normal incidence condition, where white arrows correspond to electric field vectors and the black wireframe outlines the single unit. (c) Modeled the $Q$ factor as a function of $a$ for a broken-symmetry square block ($l$=0.4 μm). Bottom inset: the schematic image of the concave Si block. The top two insets show the magnetic field distributions at around $a$=0 μm and $a$=0.4 μm, respectively. (d) Simulated $Q$ factor and resonant wavelength as a function of $\theta$ (top) and $l$ (bottom) for the BIC mode at $a$=0.4 μm. The top inset: simulated $Q$ factor as a function of $l$ at $\theta$=19°. The bottom left inset: the schematic view of the model (a rectangle block with $l_1$=0.3 μm). The bottom right inset: the magnetic field profile at $\theta$=0°. The scaling parameter is set as 1.2 in (a-d).

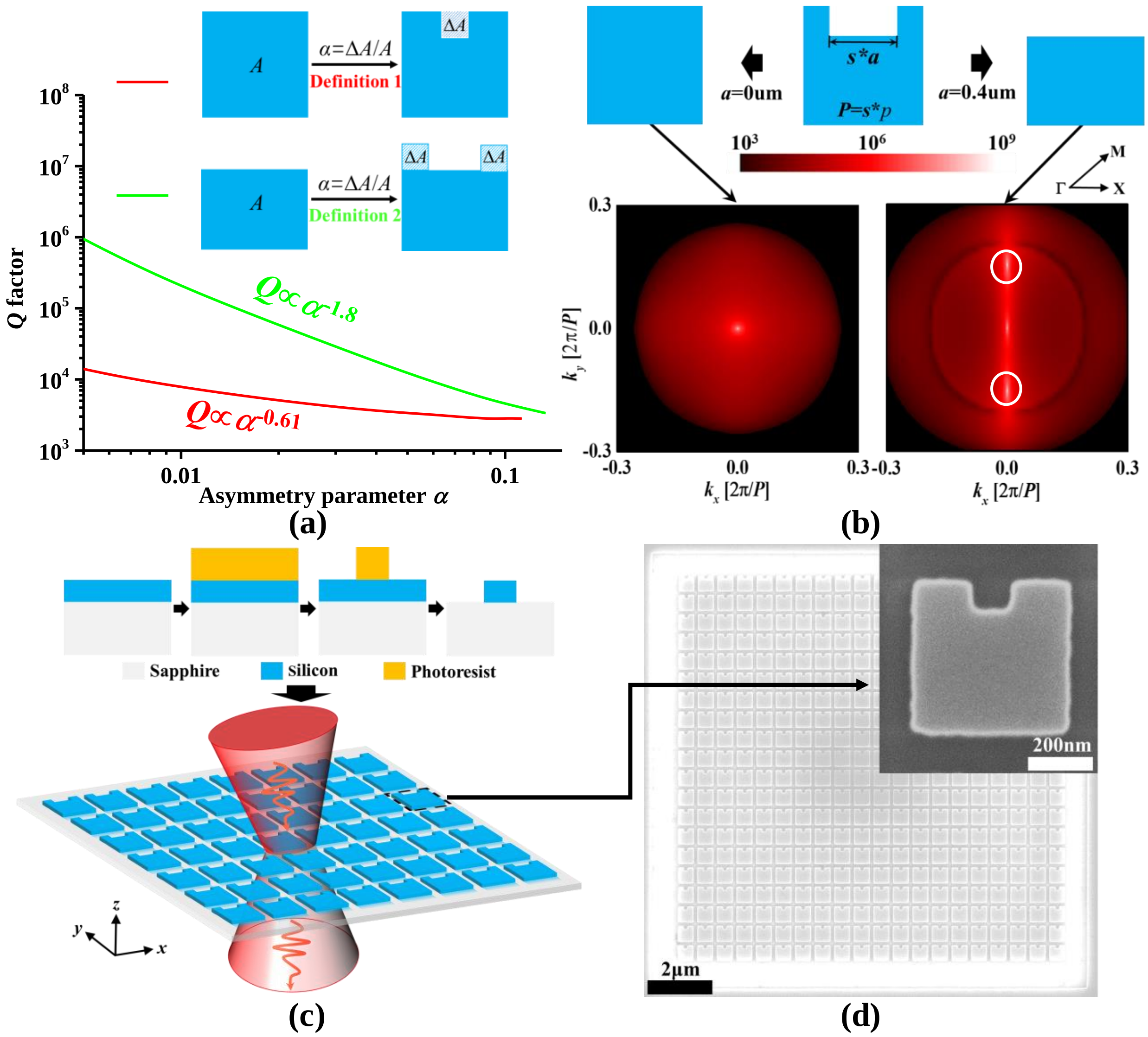

Fig. 2 | (a) Dependence of the $Q$ factor on asymmetry parameter $\alpha$ for both designs (log-log scale). Inset: two definitions of the asymmetry parameter $\alpha$. (b) Simulated $Q$ in momentum space at $a$=0 μm (left) and $a$=0.4 μm (right). Inset: the designed patterns shown in the $xy$ plane for both cases. (c) Schematic image of a period metasurface with concave nanopillars. Inset: device fabrication process. (d) SEM image of the fabricated metasurface.

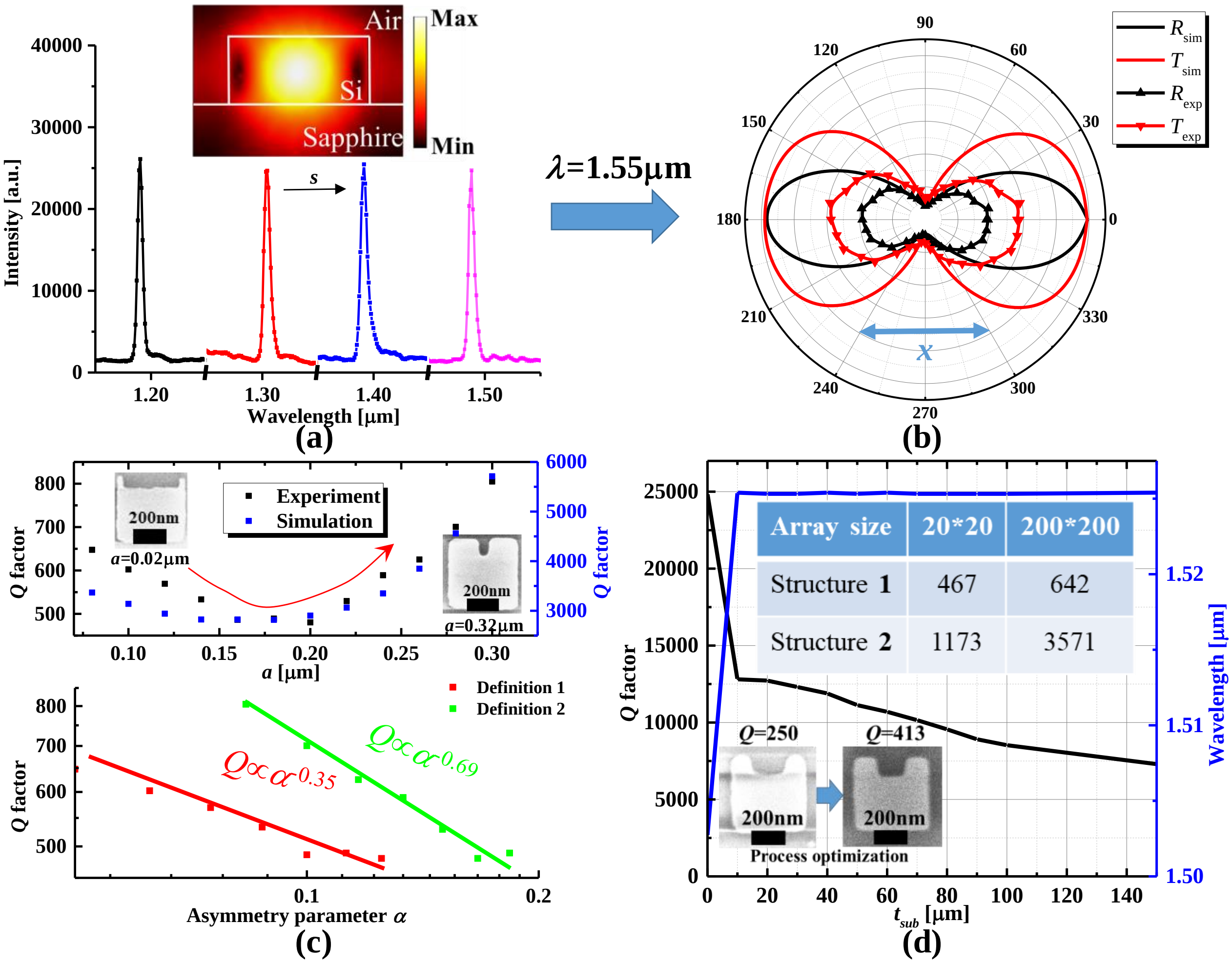

Fig. 3 | (a) Measured transmission spectrum at different $s$ (20×20 array). Inset: The magnetic field distribution in the $yz$ plane for the metasurface with the substrate. (b) Polarization dependence measurements of the transmission ($T_{exp}$) and reflection ($R_{exp}$) spectra at a wavelength of 1.55 μm (20×20 array) and their corresponding calculations (i.e. $T_{sim}$ and $R_{sim}$). The arrow indicates the direction of the $x$-axis. (c) Top: Modeled the calculated and measured $Q$ factor as a function of $a$ ($s$=1.2 and $b$=0.1 μm). Inset: SEM image of the single unit at $a$=0.02 μm and $a$=0.32 μm, respectively. Bottom: The evolution of $Q$ factor at different $\alpha$ for two definitions (log-log scale). (d) Numerically simulated dependence of the resonant wavelength and $Q$ factor on $t_{sub}$ ($t_{sub}$=0 μm means that there is no substrate). The top inset shows the comparison of the measured $Q$ factor for different dimensions and array sizes. Structure 1: $l$=0.4 μm, $a$=0.15 μm, $b$=0.1 μm, and $s$=1.15. Structure 2: $l$=0.5 μm, $a$=0.1 μm, $b$=0.1 μm, and $s$=1.02. The bottom inset shows the SEM images of the prepared structures before and after the process optimization.

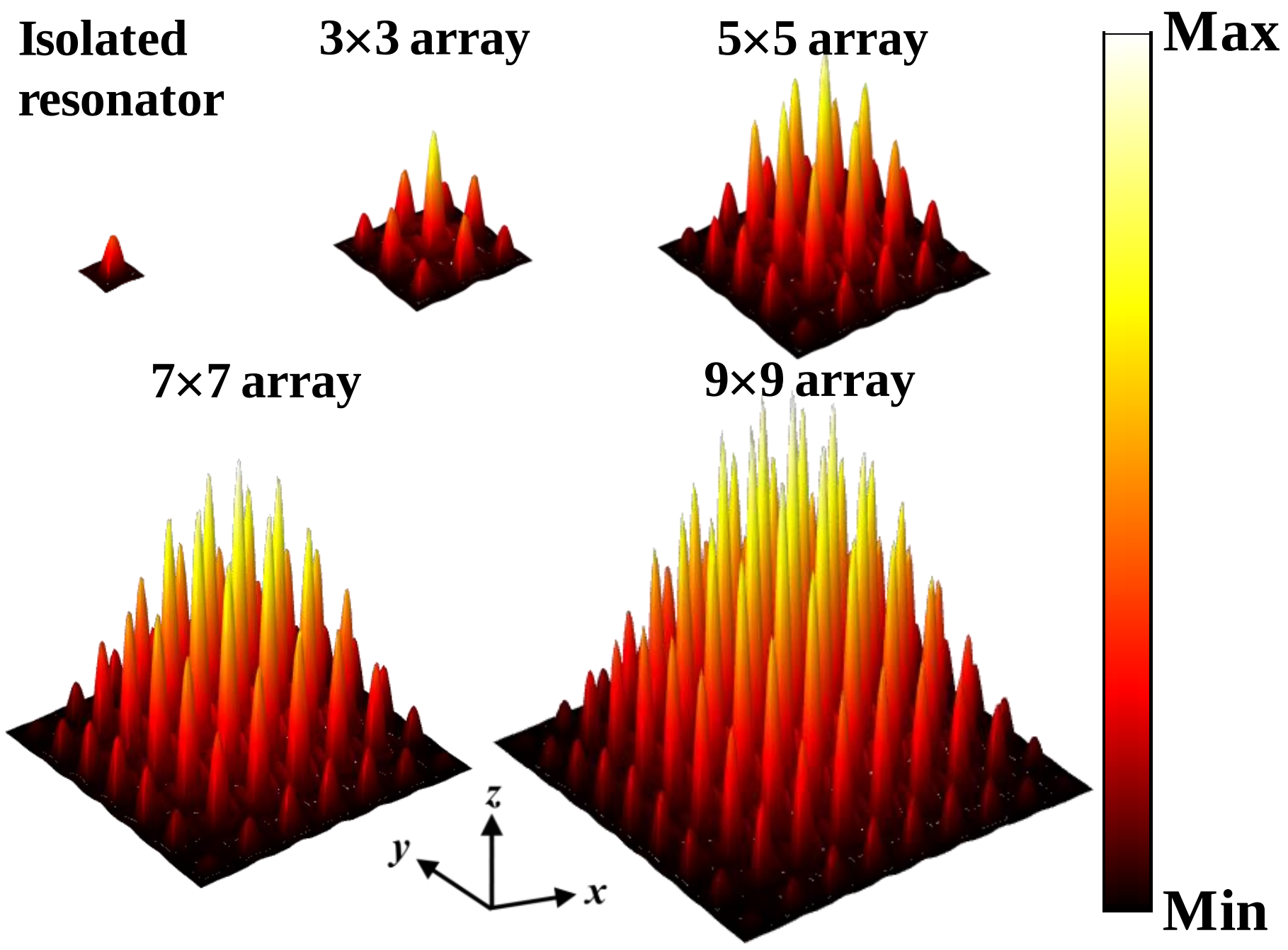

| Array size | *Q* factor |
|---|---|
| 1 | 8 |
| 3 | 66 |
| 5 | 261 |
| 7 | 605 |
| 9 | 1015 |
| infinte | 2811 |

Fig. 4 | Magnitude plots of the on-resonance magnetic field for the isolated resonator and 3×3, 5×5, 7×7, and 9×9 arrays. The magnetic field distributions are obtained at the vertical mid-plane of the blocks and the magnitude is indicated by the height. The structural parameters are $l$ =0.4 μm, $a$=0.15 μm, $b$=0.1 μm, $s$=1.2. The table on the right shows the corresponding $Q$ factors.

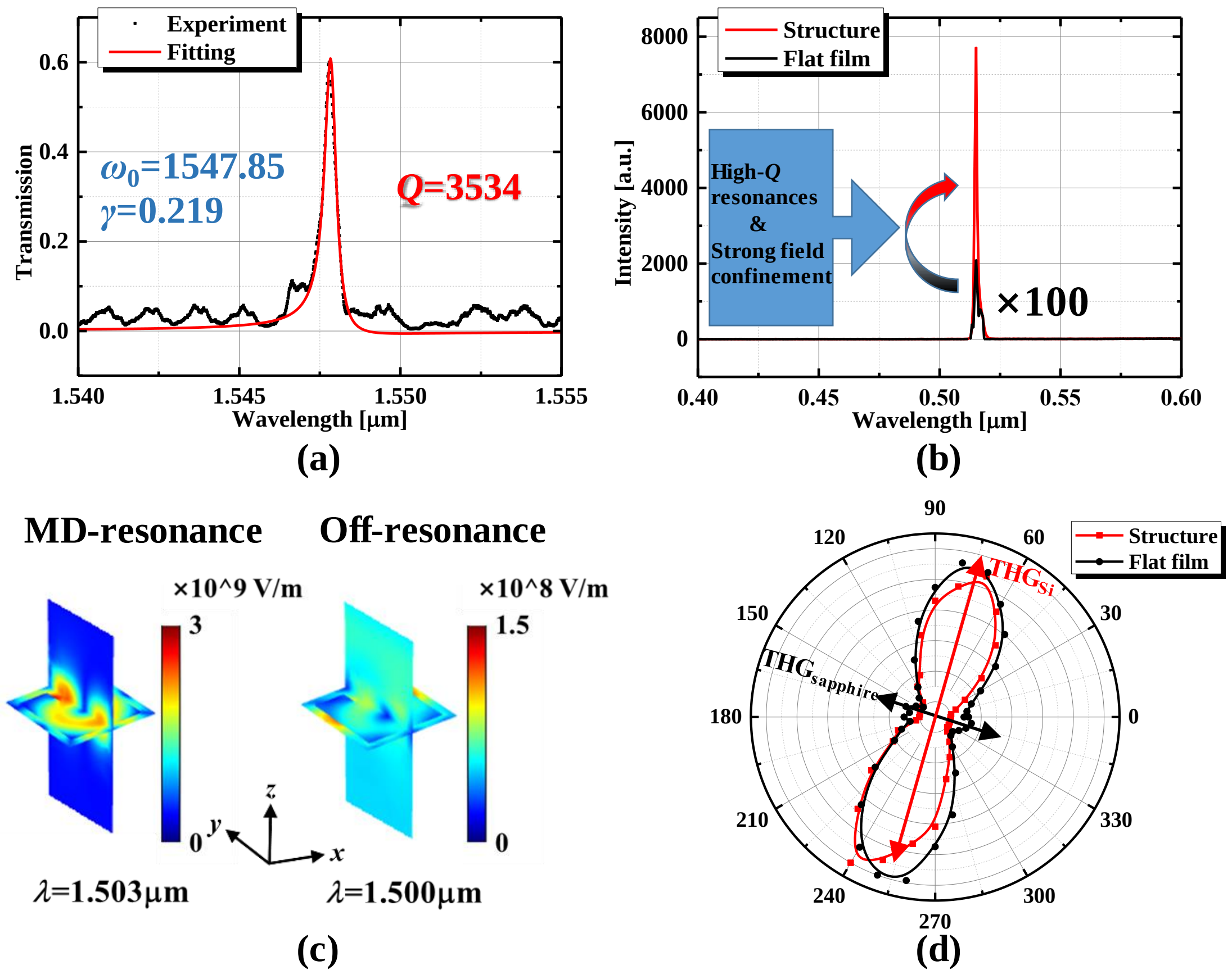

Fig. 5 | (a) Measured transmission spectrum of the proposed metasurface (200×200 array) and its fitting of Fano function. Optimized structural parameters are $p$=0.6 μm, $l$=0.5 μm, $a$=0.1 μm, $b$=0.1 μm, $s$=1.05. (b) Comparison of THG in the proposed structure (red line) and a flat Si film with the same thickness (black line). The intensity of the THG in the flat Si film is amplified 100 times. (c) The electric field profiles at the resonance (left) and the off-resonance (right), respectively. (d) The corresponding polarization dependence measurements (normalized to the peak value of the resonance).